\documentclass{article}

\usepackage{arxiv}

\usepackage[utf8]{inputenc} 
\usepackage[T1]{fontenc}    
\usepackage{hyperref}       
\usepackage{url}            
\usepackage{booktabs}       
\usepackage{amsfonts}       
\usepackage{nicefrac}       
\usepackage{microtype}      
\usepackage{lipsum}		
\usepackage{graphicx}
\usepackage{natbib}
\usepackage{doi}
\usepackage{bm}
\usepackage[varvw]{newtxmath}       
\usepackage{amsmath}
\mathchardef\mhyphen="2D 
\usepackage{booktabs}
\usepackage{multirow}
\usepackage{comment}

\title{Practical Applications of Gaussian Process with Uncertainty Quantification and Sensitivity Analysis for Digital Twin for Accident Tolerant Fuel}

\author{ \href{https://orcid.org/0000-0000-0000-0000}{\includegraphics[scale=0.06]{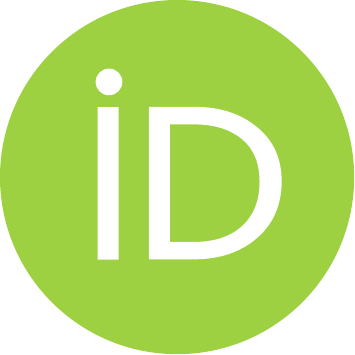}\hspace{1mm}Kazuma ~Kobayashi}\\
	Department of Nuclear Engineering and Radiation Science\\
	Missouri University of Science and Technology\\
	Rolla, MO 65409, USA \\
	\And
	\href{https://orcid.org/0000-0000-0000-0000}{\includegraphics[scale=0.06]{orcid.pdf}\hspace{1mm}Dinesh ~Kumar } \\
	Department of Mechanical Engineering\\
	University of Bristol\\
	Bristol BS8 1TR, UK \\
\And
	\href{https://orcid.org/0000-0000-0000-0000}{\includegraphics[scale=0.06]{orcid.pdf}\hspace{1mm}Matthew ~Bonney} \\
	Department of Mechanical Engineering\\
	University of Sheffield\\
	Sheffield S10 2TN, UK\\
 \And
	\href{https://orcid.org/0000-0000-0000-0000}{\includegraphics[scale=0.06]{orcid.pdf}\hspace{1mm}Syed ~Alam} \\
	Department of Nuclear Engineering and Radiation Science\\
	Missouri University of Science and Technology\\
	Rolla, MO 65409, USA \\
 }

\renewcommand{\shorttitle}{\textit{Handbook of Smart Energy Systems} }

\hypersetup{
pdftitle={A template for the arxiv style},
pdfsubject={q-bio.NC, q-bio.QM},
pdfauthor={David S.~Hippocampus, Elias D.~Striatum},
pdfkeywords={First keyword, Second keyword, More},
}

\begin{document}
\maketitle

\begin{abstract}
The application of digital twin (DT) technology to the nuclear field is one of the challenges in the future development of nuclear energy. Possible applications of DT technology in the nuclear field are expected to be very wide: operate commercial nuclear reactors, monitor spent fuel storage and disposal facilities, and develop new nuclear systems. As U.S. Nuclear Regulatory Committee (NRC) recently announced, machine learning (ML) and artificial intelligence (AI) will be new domains in the nuclear field. Considering the data science perspective, Gaussian Process (GP) has proven to be an ML algorithm for modeling and simulation components of the digital twin framework, specifically for the accident tolerant fuel (ATF) concepts. ATF is one of the high-priority areas for both the U.S. Department of Energy (DOE) and NRC. GP’s inherent treatment of lack of data, missing data, and data inconsistencies (noisy/erroneous data) present in the ATF concepts make it an attractive machine learning algorithm for implementation in the DT framework. This chapter focuses on the practical demonstration of GP and its applicability to the DT framework for predicting ATF.
\end{abstract}


\section{Introduction}
Concerns over global warming issues have attracted worldwide interest in developing sustainable clean energy technologies \citep{dell2004clean}. Nuclear power is viable for clean energy sources such as solar, hydro, and wind power because of its high energy density and immunity to weather fluctuations. However, as the accident at the Fukushima Daiichi Nuclear Power Plant has shown, the consequence of an accident in the use of nuclear power are far greater than those of other power generation methods, and rigorous safety performance assessment is required from the design stage. Against this background, the nuclear industry in recent years has focused on developing accident tolerant fuel (ATF) systems and has begun to select new nuclear fuel and cladding materials. Previous studies anticipated that ideas such as composite fuels, which are more advanced $\rm UN$ and $\rm U_{3}Si_{2}$ nuclear fuels, will be the subject of future research \citep{yang2015uo2,costa2021oxidation, Gong2020a, Gong2020b}. The concept of the ATF system, which replaces the traditional fuel and cladding material in existing light water reactors (LWRs) with accident-tolerant materials, is unlikely to change the conventional reactor core geometry significantly. However, it is not difficult to imagine that neutronics (or neutron flux distribution) and heat transfer phenomena with new materials will change. Conventional $\rm UO_{2}$ fuel and cladding materials, mainly used in commercial reactors worldwide, have a wealth of experimental data accumulated to date, and their respective properties have already been empirically tested model. Therefore, the simulation is effortless if the fuel and cladding materials are not changed. \textbf{In contrast, the case of ATF systems is not straightforward.} Since the ATF fuels were developed relatively recently, \textbf{fundamental data on their thermal and mechanical properties are scarce} \citep{Gong2020a,Gong2020b,Jaques2015,Yang2021, Ribeiro2021}. Furthermore, \textbf{only limited data are available} on the effects of irradiation under LWR operation conditions. In addition, building an empirical model requires a large amount of experimental data, and this process takes several years to complete. It is unrealistic to repeat such time-consuming and expensive experiments as many times as necessary. As a practical solution, a modeling method that makes maximum use of limited data is strongly desired in the nuclear industry. In order to build a model from very limited (experimental) data, \textbf{Kriging} is one of the promising methods; which is one of interpolation based on \textbf{Gaussian process (GP)} governed by prior covariances. Kriging delivers the best linear unbiased prediction at unsampled points assuming appropriate prior assumptions. \citep{Chung2019}. 

The advancement of simulation technology with the improvement of computer performance has enabled the development of the new field of digital twin (DT), a virtual model of a physical object that is created to reflect the physical object accurately. In addition, DT is proven to be a risk-informed and enhanced decision-making platform under multi-criteria environments for different engineering applications \citep{Pignatta2022, Ham2020}. With this technological innovation, the NRC Office of Nuclear Regulatory Research has announced a future plant to utilize it in the development of advanced nuclear systems \citep{Yadav2021, UnitedStetesNuclearRegulatoryCommision}. According to the NRC, \textbf{DT integrated with Machine Learning (ML) and Artificial Intelligence (AI)} have the potential to transform the nuclear energy sector in the coming years to help them make risk-informed decision-making, which can be leveraged for Accelerated Fuel Qualification (AFQ) for ATF \citep{Carlson2021}. 

This chapter introduces the concept of an advanced fuel development framework based on DT, and how the GP model works in the framework with example applications.

\section{Digital Twin Framework for Advanced Nuclear Fuel Development}
\begin{figure}[!htbp]
\centering
\includegraphics[scale=0.55]{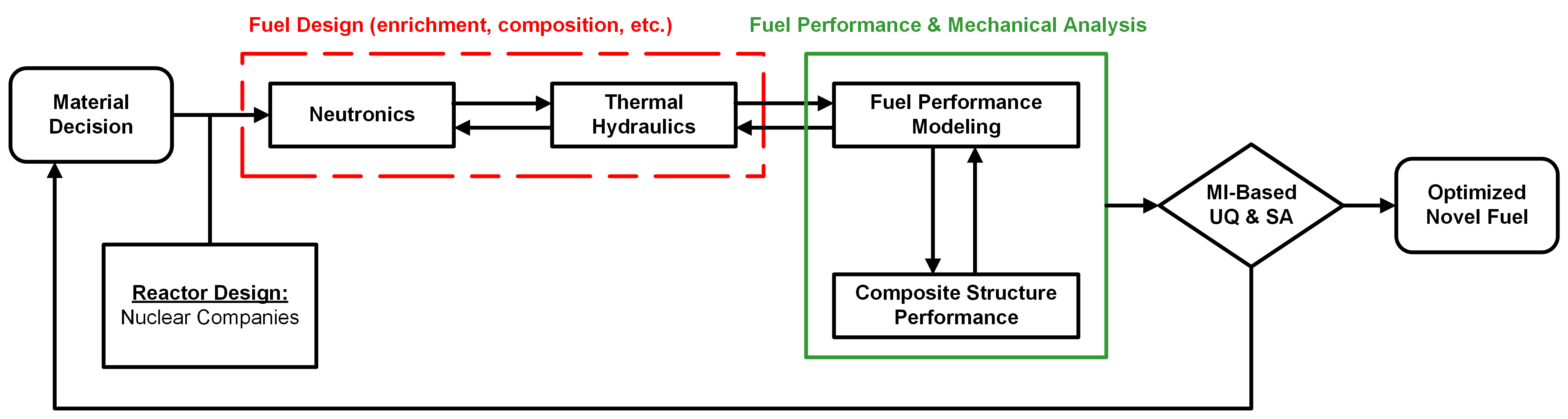}
\caption{Concept flowchart of the DT framework for ATF fuels development. The direction of the arrow indicates the execution step of each process and the data transfer. Arrows in both directions between processes represent a coupling calculation.}
\label{fig:Concept}
\end{figure}

Possible applications of DT technology in the nuclear field include the operation of commercial nuclear reactors and monitoring of spent fuel storage and disposal facilities, but the focus here is on its use in the development of ATF fuels. Fig. \ref{fig:Concept} shows the concept flowchart of the DT framework for ATF fuels development. The process shown by each block enclosed by the red dashed and green solid lines in Fig. \ref{fig:Concept} is constructed by the simulation code. MCNP \citep{lanl_mcnp}, SCALE \citep{ornl_scale}, and COBRA-NC \citep{cobra-en} are examples of codes grouped into the red dashed line. Also, BISON \citep{inl_bison} and Abaqus CAE \citep{abaqus} are the fuel performance and structure analysis codes groped into the green solid line. These code combinations are examples only, and it is expected that each university, research institute, or company will incorporate available codes as appropriate. As a reminder, arrows in both directions between processes in Fig. \ref{fig:Concept} represent a coupling calculation. Since there have been numerous studies on neutronics and thermal-hydraulic coupling simulation, please refer to the references as appropriate \citep{Alam2019, Wang2020}. Since the fuel performance code such as BISON is a relatively new development, coupling-simulation methods using this code are one of the subjects of future research.

The critical process in the proposed framework is the first section, "Material Decision." In addition to the scarcity of experimental data for newly developed fuel and cladding materials, noise during measurement is a fatal problem in creating input data (material properties). Also, due to the uncertainty of input data, it is necessary to evaluate how the uncertainties in the input data propagate errors to the results obtained in subsequent coupled simulations; therefore, uncertainty quantification (UQ) and sensitivity analysis (SA) methods must be employed. From this discussion, the following thoughts may come to mind here: \textbf{"is there any way to reduce the uncertainty of the simulation?"} A possible solution would be to replace the input data with an accurate model built using MI, such as a GP regression which is remarked as a valuable and powerful method.

\section{Gaussian Process (GP)}
The characteristics of a GP regression \citep{chakraborty2021role} are briefly summarized below:

 \begin{itemize} 
    \item Predictions from models are given in terms of probabilities rather than definite values
    \item The predicted values, also called the posterior, follow a gaussian distribution, and its confidence can be discussed over the variance
    \item The variance of the predictions will be lower if the input variables are close to the training data, in other words, increased prediction confidence 
\end{itemize}

The mathematical representation of the Gaussian process is shown in the following paragraphs of this article. In order to derive the Gaussian process, we start a discussion from a Bayesian linear regression method. In the linear regression model, the function value $f$ and observed target value $y$ are expressed as follows:
\begin{equation}
\label{eq:LR}
    f(\bm{x}) = \bm{x}^{\top}\rm \textbf{w},
\end{equation}

\begin{equation}
    y = f(\bm{x}) + \varepsilon,
\end{equation}
\noindent
where a vector of weights is defined as $\rm \textbf{w} \sim  \mathcal{N}(\bm{0},\Sigma_{p})$ and $\varepsilon$ represents a noise defined with the assumption of following the Gaussian distribution of zero mean and variance $\sigma_{n}^2$. Thus, the noise term is expressed as $\varepsilon \sim \mathcal{N}(0, \sigma_{n}^2)$.
While linear models of the input variable $\bm{x}$ have been discussed previously, nonlinear projection enables the input to be extended to higher dimensional spaces. The model can be expanded from a simple linear regression Eq. \ref{eq:LR} as below by projecting input variables to high dimensional space:

\begin{equation}
    \begin{bmatrix}
f^{(1)} \\ f^{(2)} \\ \vdots \\ f^{(n)}
    \end{bmatrix}
= \begin{bmatrix}
\phi_{0}(\bm{x}^{(1)}) && \phi_{0}(\bm{x}^{(2)}) && \cdots && \phi_{0}(\bm{x}^{(n)}) \\
\phi_{1}(\bm{x}^{(1)}) && \phi_{1}(\bm{x}^{(2)}) && \cdots && \phi_{1}(\bm{x}^{(n)}) \\
\vdots  && \vdots && && \vdots\\
\phi_{m}(\bm{x}^{(1)}) && \phi_{m}(\bm{x}^{(2)}) && \cdots && \phi_{m}(\bm{x}^{(n)}) \\
\end{bmatrix}^{\top}
\begin{bmatrix}
\rm {w_{0}} \\ \rm{w_{1}} \\ \vdots \\\rm{w_{m}}
\end{bmatrix} \Longleftrightarrow f(\bm{x}) = \bm{\phi(x)}^{\top}{\rm \textbf{w}},
\end{equation}

\noindent
where $n$ corresponds to a number of observations in a training set, $m$ is a number of desired observed values,, $\bm{\phi(x)}$ represents the mapping function.

Gaussian processes are distributions over functions $f(\bm x)$ of which the distribution is defined by a mean function $m(x)$ and the covariance function $ k(\bm{x},\bm{{x^\prime}})$ of a real function $f(\bm{x})$, where for the mean and covariance functions as

\begin{equation*}
\label{eq:mean}
    m(\bm{x}) = \mathop{\mathbb{E}}[f(\bm{x})],
\end{equation*}

\begin{equation*}
\label{eq:covariance}
    k(\bm{x},\bm{{x^\prime}}) =  \mathop{\mathbb{E}}[(f(\bm{x}) - m(\bm{x}))(f(\bm{x^{\prime}})-m(\bm{x^{\prime}}))],
\end{equation*}

\noindent
and the Gaussian process is expressed as

\begin{equation}
\label{eq:GP}
    f(\bm{x}) \sim \mathcal{GP}(m(\bm{x}),k(\bm{x},\bm{x^{\prime}})).
\end{equation}

\noindent
Note: the random variables represent the value of the function $f({x})$ at location ${x}$. The index set $X$  is the set of possible inputs, more generally for the stochastic process. Also, we define the following expression in the training set to identify the random variables. 

\noindent
For the mean and covariance can be calculated with the definition of weights $\rm \textbf{w}$,

\begin{equation}
 m(\bm{x})= \mathop{\mathbb{E}}[f(\bm{x})] = \phi(\bm{x})^{\top} \mathop{\mathbb{E}}[\rm \textbf{w}] =0,
\end{equation}

\begin{equation}
\label{eq:cov}
  k(\bm{x},\bm{x^{\prime}})= \mathop{\mathbb{E}}[f(\bm{x})f(\bm{x}^{\prime})] = \phi(\bm{x})^{\top} \mathop{\mathbb{E}}[{\rm \textbf{ww}^{\top}}]\phi(\bm{x^{\prime}}) = \phi(\bm{x})^{\top}\Sigma_{p}\phi(\bm{x}^{\prime}).
\end{equation}

\noindent
Eq. \ref{eq:cov} must be calculated to confirm distribution. There are several choices for a covariance function, $k$, the \textit{squared exponential covariance} function is the major one. The expression of this function is shown as Eq. \ref{eq:SE}.

\begin{equation}
\label{eq:SE}
    \text{cov}(f(\bm{x}_{p}),f(\bm{x}_{q})) = k(\bm{x}_{p},\bm{x}_{q}) = \text{exp}\left( -\frac{1}{2}|\bm{x}_{p} - \bm{x}_{q}  |^{2} \right),
\end{equation}
\noindent
where $\bm{x}_{p}$ and $\bm{x}_{q}$ are pairs of random variables. In another way, the covariance function can be expressed using Gram matrix, $K$,

\begin{equation}
K=
    \begin{bmatrix}
    k(\bm{x}^{(1)},\bm{x}^{(1)}) && k(\bm{x}^{(1)},\bm{x}^{(2)}) && \cdots && k(\bm{x}^{(1)},\bm{x}^{(n)})\\
    k(\bm{x}^{(2)},\bm{x}^{(1)}) && k(\bm{x}^{(2)},\bm{x}^{(2)}) && \cdots && k(\bm{x}^{(2)},\bm{x}^{(n)})\\ 
    \vdots && \vdots && && \vdots\\
    k(\bm{x}^{(n)},\bm{x}^{(1)}) && k(\bm{x}^{(n)},\bm{x}^{(2)}) && \cdots && k(\bm{x}^{(n)},\bm{x}^{(n)})\\
    \end{bmatrix}.
\end{equation}

\noindent
The GP would be used with training and test data from (noisy) experimental results. Therefore, the relationship between the distribution of the observed target (training) and the test data should be considered. We denote the training output, $\bm{f}$, and test output $\bm{f_{*}}$. For the noisy observations, the following expression appears \citep{chakraborty2021role}

\begin{equation}
\label{eq:noisy}
    \begin{bmatrix}
    \bm{y} \\ \bm{f_{*}}
    \end{bmatrix} = 
    \mathcal{N}
    \left(
    \bm{0},
    \begin{bmatrix}
    K(X,X) + \sigma_{n}^{2}I && K(X,X_{*}) \\
    K(X_{*},X) && K(X_{*},X_{*})
    \end{bmatrix}
    \right).
\end{equation}

As a result, the predictive equation for the test data can be acquired by the following equations,

\begin{equation}
    p(\bm{f}_{*}|X,\bm{y}, X_{*}) \sim \mathcal{N}(\Bar{f_{*}},\, \text{cov}(\bm{f}_{*})),
\end{equation}

\noindent
where the mean, ${\Bar{f}_{*}}$, and covariance, $\text{cov}(\bm{f}_{*})$, are given as 

\begin{equation}
    \Bar{f_{*}} = \bm{k}_{*}^{\top}(K+\sigma_{n}^{2})^{-1}\bm{y},
\end{equation}

\begin{equation}
    \mathop{\mathbb{V}}[f_{*}] = k(\bm{x}_{*}, \bm{x}_{*})-\bm{k}_{*}^{\top}(K+\sigma_{n}^{2}I)^{-1}\bm{k}_{*}.
\end{equation}

\begin{figure}[!htbp]
\centering
\includegraphics[scale=.65]{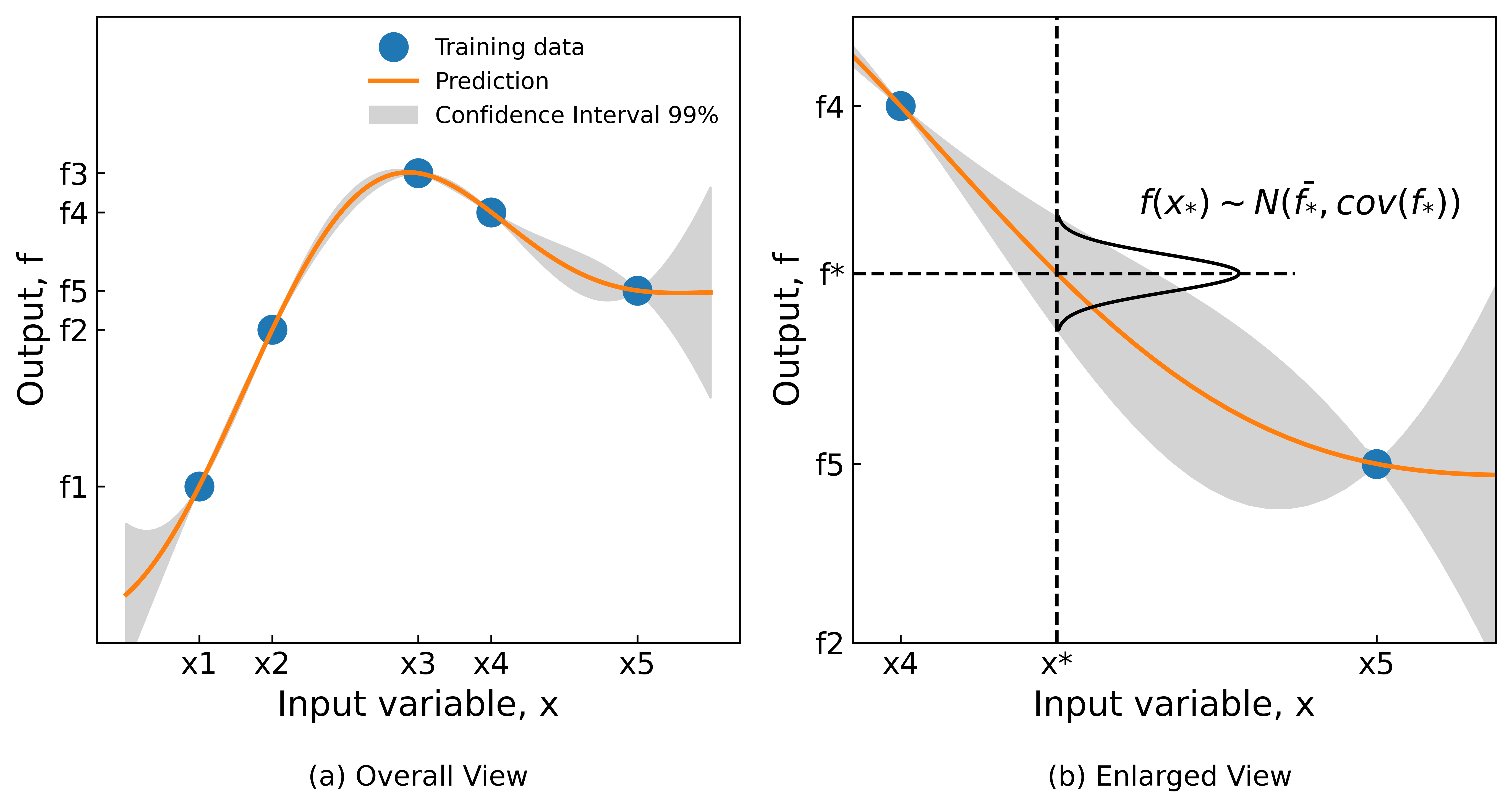}
\caption{Example of GP model to the one-dimensional dataset. Panel (a) shows the overall view of training data (dot), prediction (solid line), and the shaded area represents plus and minus three times the standard deviation for each input variable. Panel (b) is an enlarged view where the input variables are between $\rm x_4$ and $\rm x_5$. The prediction is given as a normal distribution,$\mathcal{N}(\Bar{f_{*}},  \mathop{\mathbb{V}}[f_{*}])$.}
\label{fig:GP_Sample}
\end{figure}

Fig. \ref{fig:GP_Sample} shows a sample of GP model to 1-dimension input data. The training data contains pairs of input variable and target, $(x_i, y_i)$, and represented as dot points in Fig. \ref{fig:GP_Sample}. Since the training data is shown as a scatter plot, the prediction acquired with GP can also be plotted in a 2-d plane (solid line). The shaded area corresponds to the standard deviation for each input variable plus and minus three. GP regression can be used to estimate the output value corresponding to an unknown input variable, as shown in Fig. \ref{fig:GP_Sample} (b), and the confidence of that value can be discussed further in terms of variance.

\section{Applications of a GP to Nuclear Fuel Development}
\label{sec:applications}
This section presents examples of GP model building for nuclear applications. The GP model was implemented using SMT: Surrogate Modeling Toolbox developed by the University of Michigan and its collaborators \citep{SMT2019}. A potent example of the use of this technique is the modeling of the properties of nuclear fuel systems. 

 In a nuclear reactor, the nuclear reaction rate depends on the fissile material density in the fuel, and the reaction rate significantly affects the amount of heat generated in the fuel. Therefore, evaluating the reactor's energy output and the nuclear fuel's temperature during operation is an important task when designing a reactor system employing a new type of fuel material. The problem that arises here is calculating the temperature of nuclear fuel. As a typical calculation process, neutron transport codes such as MCNP and SCALE calculate heat generation. The result is sent to a CFD code to get the temperature profile. CFD code requires the user to define thermal properties and other parameters, but those ones are usually temperature dependent and must be given as a function of temperature. Therefore, it is obvious that, aside from typical nuclear fuel such as $\rm UO_{2}$ for which experimental data and models are well equipped from past experience, it is necessary to prepare models for new materials as a pre-process. This requirement makes a GP regression the pre-processing method. 

\begin{figure}[!htbp]
\centering
\includegraphics[scale=.8]{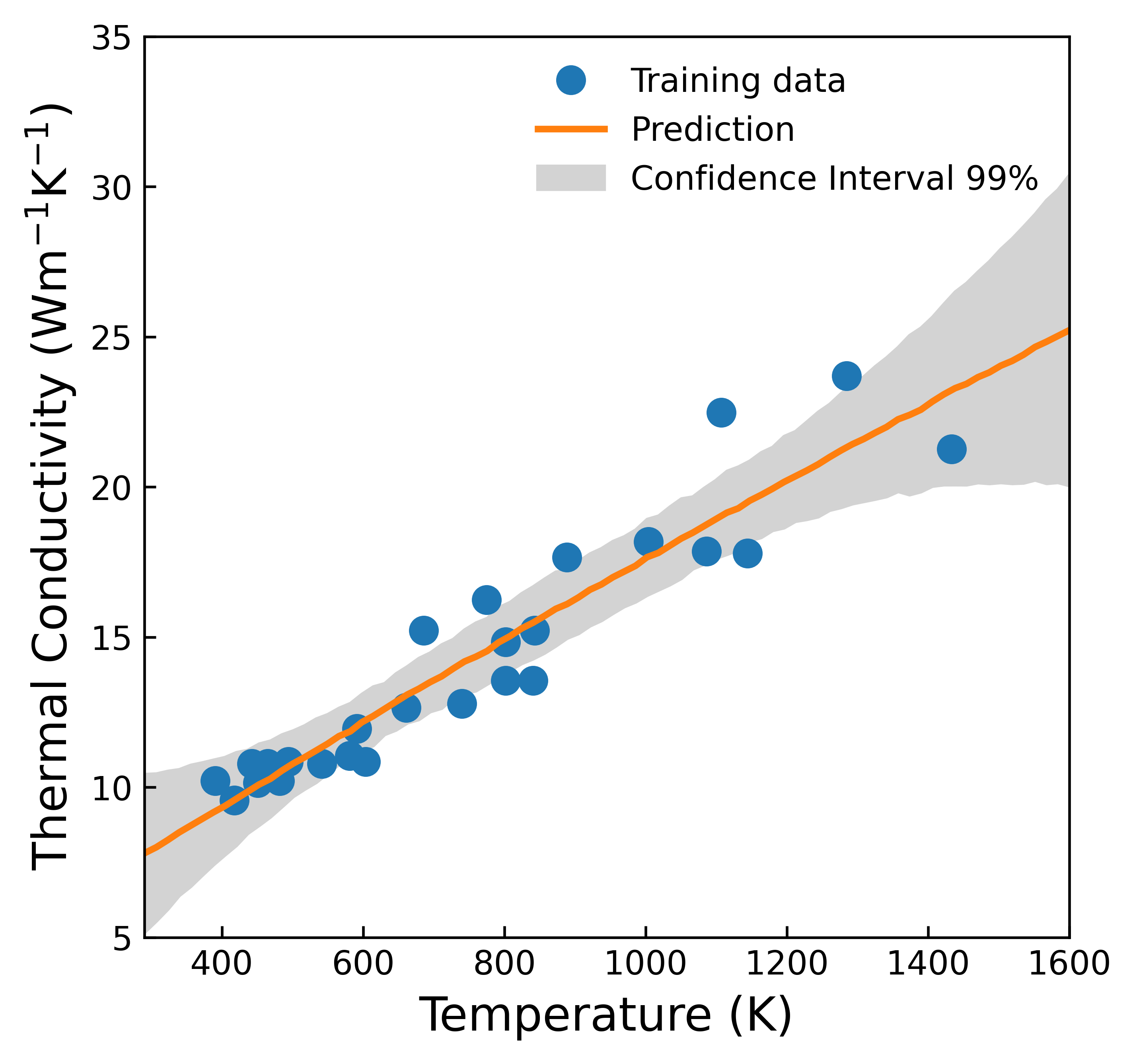}
\caption{The correlation between thermal conductivity of $\rm U_{3}Si_{2}$ and temperature from Shimizu \citep{Shimizu1965}. The dot points represent experimental data (training), the solid line the prediction by a GP, and the shaded area corresponds to the confidence interval of which is plus and minus three the standard deviation. }
\label{fig:Cond_U3Si2}
\end{figure}

$\rm U_{3}Si_{2}$ is considered to be one of the options for new nuclear fuel in terms of its superior uranium density and thermal properties. In order to build a model, experimental data reported by Shimizu \citep{Shimizu1965} was employed as training data and it is expressed as a dot plot in Fig. \ref{fig:Cond_U3Si2}. The training data contains the target value (thermal conductivity), and 1-dimensional input variable (temperature). The prediction given by a GP is the solid line and its confidence for each input value appears in Fig. \ref{fig:Cond_U3Si2}. In conventional linear regression, the optimal solution is obtained by first defining the function type and then finding its parameters. However, this method relies heavily on the function type guesses and requires skill and experience. In contrast, a GP regression can directly build a model from training data without function type estimation. Also, it is possible to discuss the validity of the values predicted by the unknown input data based on the variance values in the model.

\begin{figure}[!htbp]
\centering
\includegraphics[scale=.7]{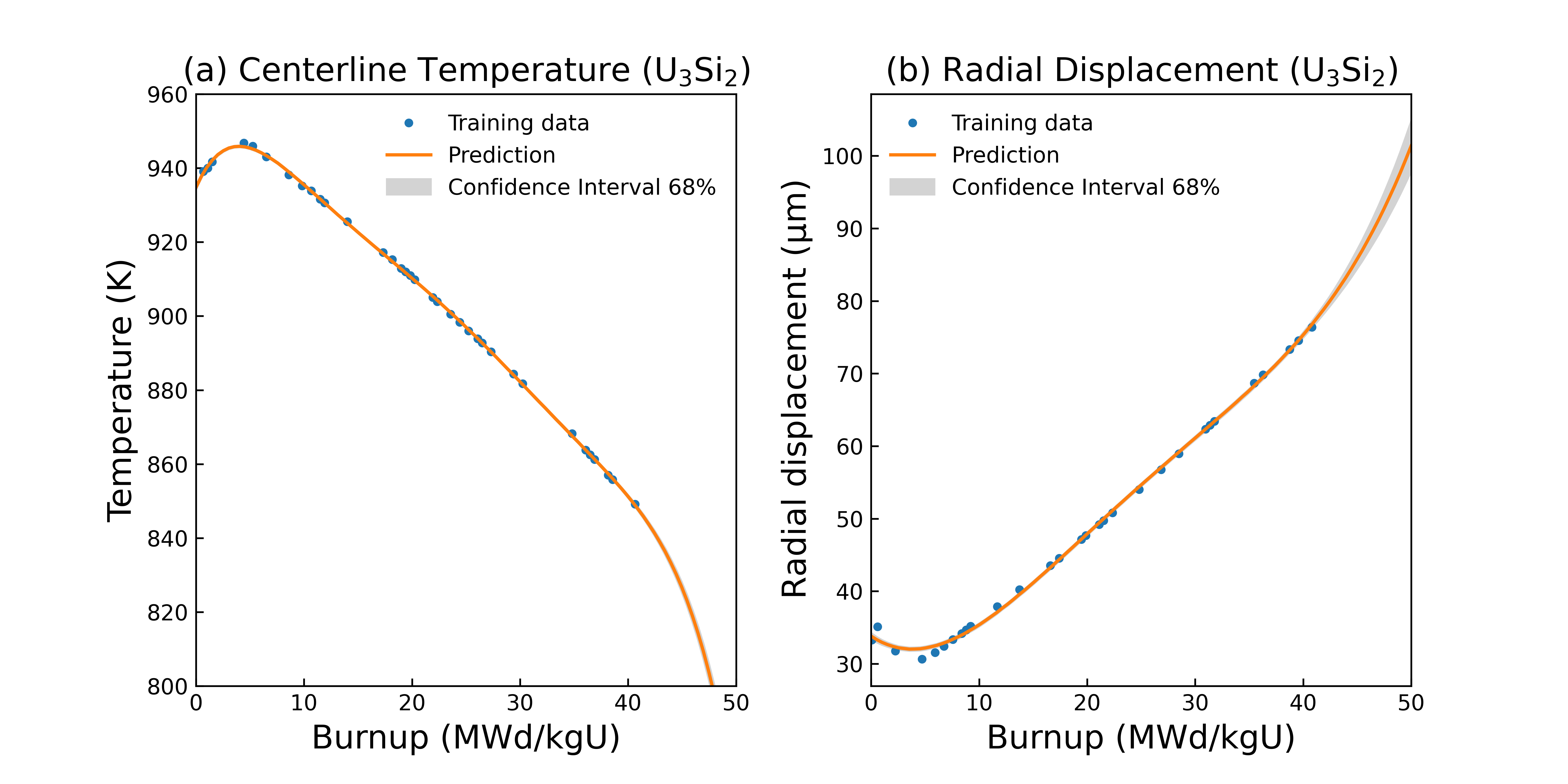}
\caption{Panel (a) represents the the correlation between centerline temperature of $\rm U_{3}Si_{2}$ fuel pellet and burnup (MWd/kgU). Panel (b) the relationship between the radial displacement and burnup (MWd/kgU): data retrieved from He et al. \citep{He2018}. The dot points represent training data, the solid lines the prediction by a GP, and the shaded areas correspond to the confidence interval which is plus and minus two the standard deviation. }
\label{fig:Temp_U3Si2}
\end{figure}

The following example is an application of GP to simulation data. Fig. \ref{fig:Temp_U3Si2} represents the fuel performance analysis conducted using a 2D axisymmetric model performed by He et al. \citep{He2018}. The plots represent the correlations between the burnup (MWd/kgU) and $\rm U_{3}Si_{2}$ fuel centerline temperature/radial displacement. Thus, modeling can be done by inputting simulation results as training data.

\begin{figure}[!htbp]
\centering
\includegraphics[scale=.7]{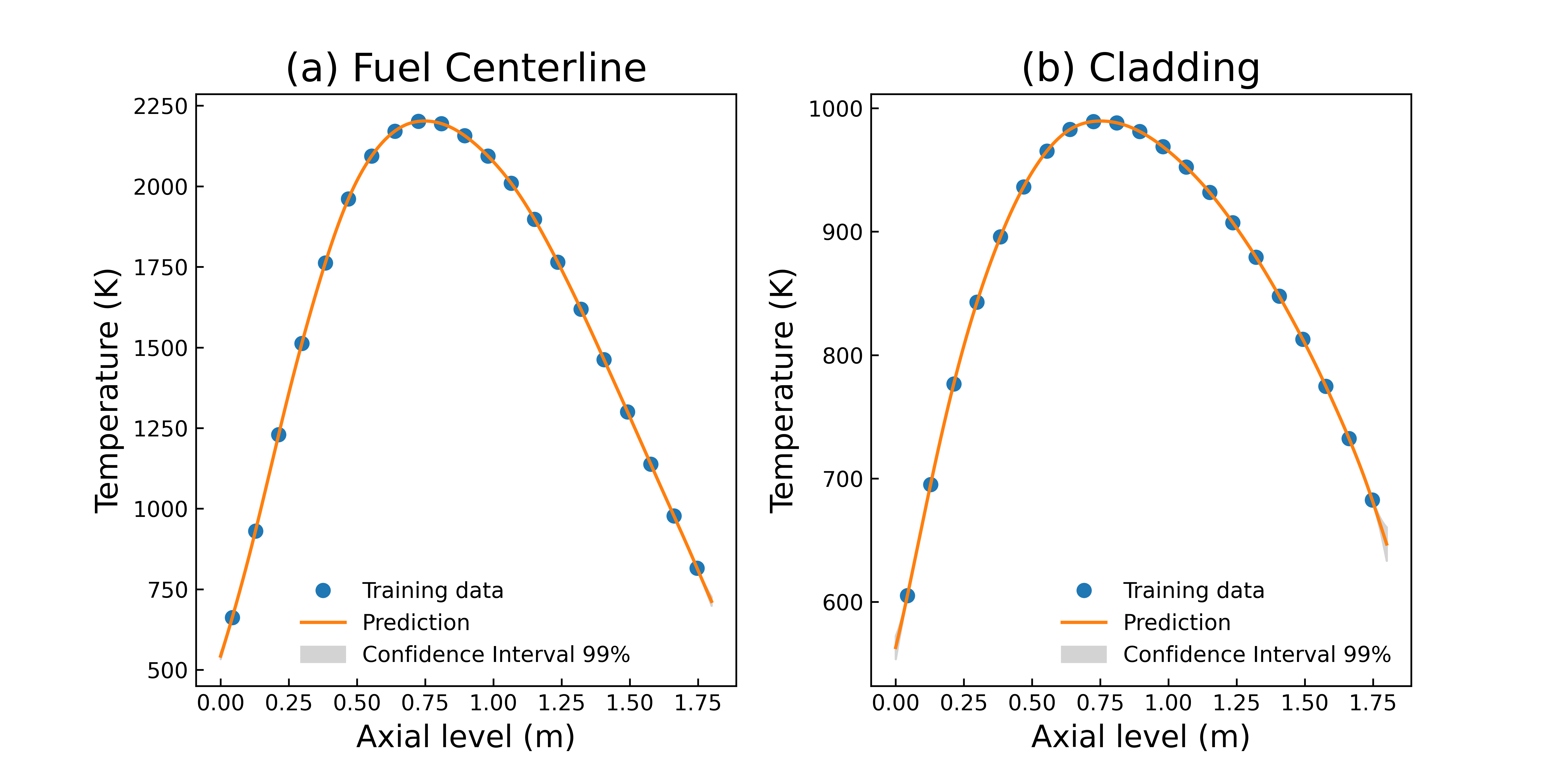}
\caption{The results of coupled neutronics and thermal hydraulic hot channel analysis of high power density civil marine SMR cores performed by S. Alam et al.   \citep{Alam2019}. Panel (a) shows the relationship between the fuel centerline temperature and axial level, and Panel (b) is for the cladding surface temperature and axial level. The fuel and cladding materials are $\rm UO_{2}$ and Zr respectively. The dot points represent training data, the solid lines the prediction by a GP, and the shaded areas correspond to the confidence interval which is plus and minus two the standard deviation.}
\label{fig:Centerline_U3Si2}
\end{figure}

 Fig. \ref{fig:Centerline_U3Si2} shows the results of coupled neutronics and thermal hydraulic hot channel analysis of high power density civil marine SMR cores performed by S. Alam et al. \citep{Alam2019}. The plots represent the correlations between the axial level of fuel and temperatures of $\rm UO_{2}$ fuel and Zr cladding.

As in the last two examples, it is possible to model using simulation results, but uncertainty assessment is insufficient. Hence, additional back-end processing such as uncertainty quantification (UQ) would be required for simulation-based modeling.

As can be seen from the example of $\rm U_{3}Si_{2}$ shown in Fig. \ref{fig:Cond_U3Si2}, outside of the range of test data, the confidence interval becomes broader. In other words, the model has a large degree of uncertainty. A simple solution to this problem is adding more training data, but one must keep the measurement limits in mind. The limit in this example is the melting point of $\rm U_{3}Si_{2}$ (1938K). Since large fluctuations in measured values can occur near the parameter at which the phase transition from solid to liquid occurs, experimental data should be obtained with sufficient caution. Based on the above discussion, when even a few additional data points are measured for thermal conductivity above 1400 K, the confidence interval after that temperature is expected to narrow.

As demonstrated in this section, GP regression can be shown to be effective in modeling nuclear fuel systems, which is a starting point in advanced nuclear system development.

\section{Conclusions}
As the U.S. NRC announced the digital twin technologies with nuclear applications, the research in this field is in its early stages, and the development potential is immense \citep{Yadav2021}. In the announcement, \textbf{handling of noisy or erroneous data} and \textbf{applications of machine learning (MI) and artificial intelligence (AI)} are listed as parts of critical challenges for future technologies. ML-based Gaussian process regression, which can build models from noisy data, could be a very powerful tool for these requirements. As it was seen in Section \ref{sec:applications}, the initial input data could be properties of nuclear fuel and other reactor components in the data-driven digital twin system for nuclear applications. Therefore, there is no question that some modeling method needs to be implemented in the system. 

The advantage of Gaussian process regression has already been mentioned in Section \ref{sec:applications} as the ability to construct models directly from training data. In addition, it is easily extended to multidimensional input variables. As an example, a model of heat conduction for $\rm U_{3}Si_{2}$ was constructed assuming that the input variable was only temperature. However, GP regression allows for more complex models to be built by increasing the type of that input variable. This advantage is expected to be utilized in the modeling of composite fuels, which have received much attention as ATF candidates, such as $\rm UO_{2}\mhyphen UN$ and $\rm UO_{2}\mhyphen U_{3}Si_{2}$ composites. These composite materials are a new type of nuclear fuel developed to overcome the weak points of $\rm UN$ and $\rm U_{3}Si_{2}$. However, it has been reported that fabrication parameters (the content of $\rm UN$ and $\rm U_{3}Si_{2}$, sintering temperature, cooling time, etc.) affect the material density and thermal properties \citep{Jaques2015,Gong2020a, Ribeiro2021}. In this case, that information can be used for building a model by treating the information as multi-dimensional input variables. Table \ref{tab:dataset} is a layout of the training data set for multi-dimensional variables; the "Sample" label represents the number of output and input variables pairs, and the "output" is a target value for corresponding input variables. Superscripts on output value and input variables correspond to the sample data set. Subscripts on input variables represent independent variables, e.g., temperature, uranium enrichment, grain size, etc. The modeling can be done by providing a multi-dimensional vectors, $  \bm{x}^{(i)} = (x_{0}^{(i)}, x_{1}^{(i)}, \cdots x_{n}^{(i)})$, as input. 

\begin{table}[]
\centering
\caption{Sample layout of training data for multi-dimensional input variables}
\label{tab:dataset}
\begin{tabular}{@{}clcccccclc@{}}
\toprule
\multicolumn{2}{c}{\multirow{2}{*}{\textbf{Sample}}} & \textbf{} & \multirow{2}{*}{\textbf{Output}} & \textbf{} & \multicolumn{5}{c}{\textbf{Input variable (s)}} \\ \cmidrule(l){6-10} 
\multicolumn{2}{c}{}         & \textbf{} &           & \textbf{} & $x_{0}$       & $x_{1}$       & \multicolumn{2}{c}{$\cdots$} & $x_{n}$       \\ \midrule
\multicolumn{2}{c}{1}        &           & $y^{(1)}$ &           & $x_{0}^{(1)}$ & $x_{1}^{(1)}$ & \multicolumn{2}{c}{$\cdots$} & $x_{n}^{(1)}$ \\
\multicolumn{2}{c}{2}        &           & $y^{(2)}$ &           & $x_{0}^{(2)}$ & $x_{1}^{(2)}$ & \multicolumn{2}{c}{$\cdots$} & $x_{n}^{(2)}$ \\
\multicolumn{2}{c}{$\vdots$} &           & $\vdots$  &           & $\vdots$      &               & \multicolumn{2}{c}{}         & $\vdots$      \\
\multicolumn{2}{c}{N}        &           & $y^{(N)}$ &           & $x_{0}^{(N)}$ & $x_{1}^{(N)}$ & \multicolumn{2}{c}{$\cdots$} & $x_{n}^{(N)}$ \\ \bottomrule
\end{tabular}
\end{table}
In the development of nuclear materials, many items must be verified, and experimental data may be insufficient even with Gaussian processes. One possible solution to address this issue is a multi-fidelity modeling method where both high-fidelity (e.g., experiment) and low-fidelity (e.g., physics model) data can be employed \citep{Chakraborty2021}. In other situations, even the physical model may be incomplete \citep{Brunton2016}. For such a worst case, a multi-fidelity physics-informed deep neural network (MF-PIDNN) is proposed \citep{Chakraborty2021}. As these examples show, despite the rapid development in the sophistication of data analysis methods, their application to the nuclear field is currently minimal. 

By actively incorporating these MI-based technologies, future nuclear technologies have the potential to gain higher reliability and further accelerate research and development.

\section*{Acknowledgement}
The computational part of this work was supported in part by the National Science Foundation (NSF) under Grant No. OAC-1919789.

\bibliographystyle{unsrtnat}
\bibliography{references}  






\end{document}